\documentclass[11pt]{article}
\usepackage{latexsym,amssymb,amsmath}
\begin{document}
\baselineskip=16pt

\newcommand{\la}{\langle}
\newcommand{\ra}{\rangle}
\newcommand{\psp}{\vspace{0.4cm}}
\newcommand{\pse}{\vspace{0.2cm}}
\newcommand{\ptl}{\partial}
\newcommand{\dlt}{\delta}
\newcommand{\sgm}{\sigma}
\newcommand{\al}{\alpha}
\newcommand{\be}{\beta}
\newcommand{\G}{\Gamma}
\newcommand{\gm}{\gamma}
\newcommand{\vs}{\varsigma}
\newcommand{\Lmd}{\Lambda}
\newcommand{\lmd}{\lambda}
\newcommand{\td}{\tilde}
\newcommand{\vf}{\varphi}
\newcommand{\yt}{Y^{\nu}}
\newcommand{\wt}{\mbox{wt}\:}
\newcommand{\rd}{\mbox{Res}}
\newcommand{\ad}{\mbox{ad}}
\newcommand{\stl}{\stackrel}
\newcommand{\ol}{\overline}
\newcommand{\ul}{\underline}
\newcommand{\es}{\epsilon}
\newcommand{\dmd}{\diamond}
\newcommand{\clt}{\clubsuit}
\newcommand{\vt}{\vartheta}
\newcommand{\ves}{\varepsilon}
\newcommand{\dg}{\dagger}
\newcommand{\tr}{\mbox{Tr}}
\newcommand{\ga}{{\cal G}({\cal A})}
\newcommand{\hga}{\hat{\cal G}({\cal A})}
\newcommand{\Edo}{\mbox{End}\:}
\newcommand{\for}{\mbox{for}}
\newcommand{\kn}{\mbox{ker}}
\newcommand{\Dlt}{\Delta}
\newcommand{\rad}{\mbox{Rad}}
\newcommand{\rta}{\rightarrow}
\newcommand{\mbb}{\mathbb}
\newcommand{\lra}{\Longrightarrow}

\begin{center}{\Large \bf Algebraic Approaches to the
Geopotential}\end{center}\begin{center}{\Large \bf Forecast  and
Nonlinear MHD Equations}\footnote {2000 Mathematical Subject
Classification. Primary 35C05, 35Q35; Secondary 35C10, 35C15.}
\end{center}
\vspace{0.2cm}

\begin{center}{\large Xiaoping Xu}\end{center}
\begin{center}{Institute of Mathematics, Academy of Mathematics \& System Sciences}\end{center}
\begin{center}{Chinese Academy of Sciences, Beijing 100190, P.R. China}
\footnote{Research supported
 by China NSF 10871193}\end{center}

\vspace{0.6cm}

 \begin{center}{\Large\bf Abstract}\end{center}

\vspace{1cm} {\small In this paper, we use various anstazes
motivated from our earlier works on transonic gas flows, boundary
layer problems and Navier-Stokes equations to find new explicit
exact solutions with multiple parameter functions for the equation
of geopotential forecast and the equations of nonlinear
magnetohydrodynamics.}

\section{Introduction}

Partial differential equation
$$(H_{xx}+H_{yy})_t+H_x(H_{xx}+H_{yy})_y-H_y(H_{xx}+H_{yy})_x=k
H_x\eqno(1.1)$$ is used in earth sciences for geopotential
forecast on a middle level (e.g., cf [Ki] and Page 222 in [I]),
where $k$ is a real constant. Kibel' [Ki] found the Gaurvitz
solution of the above equation. The well known Syono solution was
given in [Sy]. Katkov [Ka1, Ka2] determined the Lie point
symmetries and obtained certain invariant solutions. The other
known solutions are related to the physical backgrounds such as
configuration of type of narrow gullies and crests, flows of type
of isolate whirlwinds, stream flow, springs and drains, hyperbolic
points, and cyclone formation (e.g., cf. Pages 225, 226 in [I]).

In magnetohydrodynamics, it is very important to study the
nonlinear MHD equations:
$$\psi_t+\vf_x\psi_y-\vf_y\psi_x=\vf_z,\eqno(1.2)$$
\begin{eqnarray*}\hspace{2.2cm}& &(\vf_{xx}+\vf_{yy})_t+
\vf_x(\vf_{xx}+\vf_{yy})_y-\vf_y(\vf_{xx}+\vf_{yy})_x\\ &=&
(\psi_{xx}+\psi_{yy})_z+
\psi_x(\psi_{xx}+\psi_{yy})_y-\psi_y(\psi_{xx}+\psi_{yy})_x,
\hspace{3.7cm}(1.3)\end{eqnarray*} where $\vf$ and $\psi$ are the
potentials for the velocity and the transverse component of the
magnetic field, respectively (they can also be interpreted as the
potential of an electric field and the $z$-component of the vector
potential of the magnetic field) (e.g., cf. [KP] and Page 390 in
[I]). We refer [P] for more information on magnetohydrodynamics.
Samokhin [S] (1985) determined the Lie point symmetries of the above
equations, conservation laws and some solutions in terms the
solutions of the other partial differential equations. Bershadskii
[B] found a connection between the energy conservation law and the
uniqueness of classical solution of the nonlinear MHD equations. The
reason of solving the geopotential equation together with the
nonlinear MHD equations is that the left hand side of the equation
(1.1) coincides with that of the equation (1.3).

Based our earlier works on transonic gas flows [X1], boundary layer
problems [X2] and  Navier-Stokes equations [X3], we give in this
paper various ansatzes related to algebraic characteristics of the
above equations to find new explicit exact solutions with multiple
parameter functions. By specifying these parameter functions, one
can obtain the solutions of certain initial-value problems of the
above equations.

The symmetry group of the geopotential equation (1.1) is generated
by the following transformations:
$$T_{a,b}(H)=H(t+a,x,y+b),\qquad T_c(H)=c^3H(c^{-1}t,cx,cy),\qquad
a,b,c\in\mbb{R},\;c\neq 0;\eqno(1.4)$$
$$T_{\al,\be}(H)=H(t,x+\al,y)+\al'y+\be,\eqno(1.5)$$
where $\al$ and $\be$ are arbitrary functions of $t$. The symmetry
transformations of the nonlinear MHD equations (1.2) and (1.3)
that we are concerned with are:
$$T_{a,b}(\vf)=\vf(t+a,x,y,z+b),\qquad
 T_{a,b}(\psi)=\psi(t+a,x,y,z+b),\qquad a,b\in\mbb{R};\eqno(1.6)$$
$$T_{1c}(\vf)=c^{-1}\vf(ct,x,y,cz),\qquad
T_{1c}(\psi)=c^{-1}\psi(ct,x,y,cz),\eqno(1.7)$$
$$T_{2c}(\vf)=c^{-2}\vf(t,cx,xy,c),\qquad
T_{2c}(\psi)=c^{-2}\psi(t,cx,cy,z)\eqno(1.8)$$ for $0\neq
c\in\mbb{R};$
$$T_{\sgm,\tau}(\vf)=\vf(t,x+\sgm,y,z)+\sgm_t y+\tau_t,
\qquad T_{\sgm,\tau}(\psi)=\psi(t,x+\sgm,y,z)+\sgm_z
y+\tau_z,\eqno(1.9)$$
$$T_\sgm(\vf)=\vf(t,x,y+\sgm,z)-\sgm_t x,
\qquad T_\sgm(\psi)=\psi(t,x,y+\sgm,z)-\sgm_z x,\eqno(1.10)$$
where $\sgm$ and $\tau$ are any functions of $t,z$;
$$T_\al(\vf)=\vf(t,x\cos2\al+y\sin2\al,-x\sin2\al+y\cos2\al,z)
+\al'(x^2+y^2),\eqno(1.11)$$
$$T_\al(\psi)=\psi(t,x\cos2\al+y\sin2\al,-x\sin2\al+y\cos2\al,z)
+\al'(x^2+y^2),\eqno(1.12)$$
$$T_\be(\vf)=\vf(t,x\cos2\be+y\sin2\be,-x\sin2\be+y\cos2\be,z)
+\be'(x^2+y^2),\eqno(1.13)$$
$$T_\be(\psi)=\psi(t,x\cos2\be+y\sin2\be,-x\sin2\be+y\cos2\be,z)
-\be'(x^2+y^2),\eqno(1.16)$$ where $\al=\al(t+z)$ and
$\be=\be(t-z)$ are arbitrary one-variable functions. The above
transformations change solutions to solutions. For convenience, we
always assume that all the involved partial derivatives of related
functions always exist and we can change orders of taking partial
derivatives. We also use prime $'$ to denote the derivative of any
one-variable function.

In Section 2, we solve the equation (1.1) of geopotential
forecast. We find explicit exact solutions of the nonlinear MHD
equations (1.2) and (1.3) in Section 3.

\section{Solutions of Geopotential Forecast Equation}

In this section, we find two families of exact solutions of the
geopotential forecast equation (1.1).

Let $\al$ and $\be$ be functions of $t$. Set
$$\varpi=\al x+\be y.\eqno(2.1)$$
Assume
$$H=\phi(t,\varpi)+\mu y^2+\tau x+\nu
y,\eqno(2.2)$$
\newpage

\noindent where $\phi$ is a two-variable function and
$\tau,\mu,\nu$ are functions in $t$. Note
$$H_x=\al\phi_\varpi+\tau,\qquad H_y=\be\phi_\varpi+2\mu
y+\nu,\qquad
H_{xx}+H_{yy}=2\mu+(\al^2+\be^2)\phi_{\varpi\varpi},\eqno(2.3)$$
$$
(H_{xx}+H_{yy})_t=2\mu'+(\al^2+\be^2)'\phi_{\varpi\varpi}+
(\al^2+\be^2)[\phi_{t\varpi\varpi}+(\al'x+\be'y)\phi_{\varpi\varpi\varpi}],\eqno(2.4)$$
$$(H_{xx}+H_{yy})_x=(\al^2+\be^2)\al\phi_{\varpi\varpi\varpi},
\qquad(H_{xx}+H_{yy})_y=(\al^2+\be^2)\be\phi_{\varpi\varpi\varpi}.
\eqno(2.5)$$ Thus (1.1) becomes
\begin{eqnarray*}\hspace{2cm} & &2\mu'+(\al^2+\be^2)'\phi_{\varpi\varpi}+
(\al^2+\be^2)\phi_{t\varpi\varpi}-k(\al\phi_\varpi+\tau)\\ & & +
(\al^2+\be^2)[\al'x+(\be'-2\al\mu)y+\be\tau-\al\nu]
\phi_{\varpi\varpi\varpi}=0.\hspace{3.7cm}(2.6)\end{eqnarray*}

In order to solve the above equation, we assume
$$2\mu'=k\tau,\qquad \tau=\al{\vt'}',\qquad\nu=
\be{\vt'}',\eqno(2.7)$$ for some function $\vt$ of $t$, and
$$\al'x+(\be'-2\al\mu)y=0.\eqno(2.8)$$
Note that (2.8) is equivalent to the following system of ordinary
differential equations:
$$\al'=0,\qquad \be'-2\al\mu=0.\eqno(2.9)$$

By the first equation in (2.9), we have $\al=c\in\mbb{R}.$ So
$\tau=c{\vt'}'$ according to the second equation in (2.7). Moreover,
 the first equation in (2.7) yield
$$ \mu=\frac{kc\vt'+c_0}{2},\qquad c_0\in\mbb{R}.\eqno(2.10)$$
Hence the second equation in (2.9) becomes
$$\be'-c(kc\vt'+c_0)=0.\eqno(2.11)$$
Therefore,
$$\be=c(kc\vt+c_0t)+d,\qquad d\in\mbb{R}.\eqno(2.12)$$
According to the third equation in (2.7),
$$\nu=[c(kc\vt+c_0t)+d]{\vt'}'.\eqno(2.13)$$

 Now (2.6) becomes
$$(\al^2+\be^2)'\phi_{\varpi\varpi}+
(\al^2+\be^2)\phi_{t\varpi\varpi}-kc e^\gm\phi_\varpi=0.
\eqno(2.14)$$ Modulo the transformation in (1.5), it is enough to
solve the following equation:
$$(\al^2+\be^2)'\phi_\varpi+
(\al^2+\be^2)\phi_{t\varpi} -kc e^\gm\phi=0. \eqno(2.15)$$ The above
equation can written as
$$[(\al^2+\be^2)\phi_\varpi]_t-kc\phi=0.\eqno(2.16)$$
So we take the form
$$\phi=\frac{\hat\phi(t,\varpi)}{\al^2+\be^2}=\frac{\hat\phi(t,\varpi)}
{c^2+[c(kc\vt+c_0t)+d]^2}.\eqno(2.17)$$ Then (2.16) becomes
$$\hat\phi_{\varpi t}=\frac{kc\hat\phi}{c^2+[c(kc\vt+c_0t)+d]^2}.\eqno(2.18)$$
Thus we have the solution:
\begin{eqnarray*}\hspace{2cm}\hat\phi&=&\sum_{i=1}^md_i\exp\left(\frac{kca_i}{a_i^2+b_i^2}
\int\frac{dt}{c^2+[c(kc\vt+c_0t)+d]^2}+a_i\varpi\right)\\ &
&\times\sin \left(b_i\varpi+c_i-\frac{kcb_i}{a_i^2+b_i^2}
\int\frac{dt}{c^2+d^2e^{4kc^2\vt}}\right),\hspace{4.3cm}
(2.19)\end{eqnarray*} where $a_i,b_i,c_i,d_i$ are real constants
such that $(a_i,b_i)\neq (0,0)$.\psp

{\bf Theorem 2.1}. {\it Let $\vt$ be any function of $t$ and let
$a_i,b_i,c_i,d_i,c_0,c,d$ for $i=1,...,m$ be real constants such
that $(c,d),(a_i,b_i)\neq (0,0)$. We have the following solution of
the geopotential forecast equation (1.1):
\begin{eqnarray*}
H&=&\frac{kc\vt'+c_0}{2}y^2+{\vt'}'[cx+[c(kc\vt+c_0t)+d]y]+
\frac{1} {c^2+[c(kc\vt+c_0t)+d]^2}\\
& &\times\sum_{i=1}^md_i\exp\left(\frac{kca_i}{a_i^2+b_i^2}
\int\frac{dt}{c^2+[c(kc\vt+c_0t)+d]^2}+a_i[cx+[c(kc\vt+c_0t)+d]y]\right)\\
& &\times\sin
\left(b_i[cx+[c(kc\vt+c_0t)+d]y]+c_i-\frac{kcb_i}{a_i^2+b_i^2}
\int\frac{dt}{c^2+[c(kc\vt+c_0t)+d]^2}\right).\hspace{0.6cm}
(2.20)\end{eqnarray*}}

Next we set
$$\varpi=x^2+y^2.\eqno(2.21)$$
Assume
$$H=\xi(\varpi)-y\eqno(2.22)$$
where  $\xi$ is a one-variable function. Note
$$H_x=2x\xi',\qquad H_y=2y\xi'-1,\qquad
H_{xx}+H_{yy}=4(\xi'+\varpi{\xi'}'),\eqno(2.23)$$
$$(H_{xx}+H_{yy})_x=8x(2{\xi'}'+\varpi{{\xi'}'}'),\qquad
(H_{xx}+H_{yy})_y=8y(2{\xi'}'+\varpi{{\xi'}'}').\eqno(2.24)$$ Then
(1.1) is equivalent to:
$$4(2{\xi'}'+\varpi{{\xi'}'}')=k\xi'.\eqno(2.25)$$
Modulo the transformation (1.5), we only need to solve the equation:
$$\xi'+\varpi{\xi'}'=\frac{k}{4}\xi.\eqno(2.26)$$
To solve the above ordinary differential equation, we assume
$$\xi=\sum_{i=0}^\infty \varpi^i(a_i+b_i\ln\varpi),\qquad
a_i,b_i\in\mbb{R}.\eqno(2.27)$$ Observe
$$\xi'=\sum_{i=0}^\infty
\varpi^{i-1}(ia_i+b_i+ib_i\ln\varpi),\eqno(2.28)$$
$${\xi'}'=\sum_{i=0}^\infty
\varpi^{i-2}(i(i-1)a_i+(2i-1)b_i+i(i-1)b_i\ln\varpi).\eqno(2.29)$$
So (2.26) becomes
$$\sum_{i=0}^\infty
\varpi^{i-1}(i^2a_i+2ib_i+i^2b_i\ln\varpi)=\frac{k}{4}\sum_{i=0}^\infty
\varpi^i(a_i+b_i\ln\varpi),\eqno(2.30)$$ equivalently,
$$(i+1)^2a_{i+1}+2(i+1)b_{i+1}=\frac{k}{4}a_i,\qquad
(i+1)^2b_{i+1}=\frac{k}{4}b_i.\eqno(2.31)$$ Hence
$$b_i=\frac{b_0k^i}{(i!)^24^i},\qquad
a_i=\frac{a_0k^i}{(i!)^24^i}-2b_0\sum_{r=0}^i\frac{k^r}{r(r!)^24^r}.
\eqno(2.32)$$ Thus
$$\xi=a_0\sum_{i=0}^{\infty}\frac{k^i\varpi^i}{(i!)^24^i}
+b_0\sum_{j=0}^{\infty}\varpi^j
\left(\frac{k^j\ln\varpi}{(j!)^24^j}-2\sum_{r=0}^j\frac{k^r}{r(r!)^24^r}\right).
\eqno(2.33)$$\pse

{\bf Theorem 2.2}. {\it Let $b$ and $c$ be any real constants. We
have the following steady solution of the geopotential forecast
equation (1.1):
$$H=-y+b\sum_{i=0}^{\infty}\frac{k^i(x^2+y^2)^i}{(i!)^24^i}
+c\sum_{j=0}^{\infty}(x^2+y^2)^j
\left(\frac{k^j\ln(x^2+y^2)}{(j!)^24^j}-2\sum_{r=0}^j\frac{k^r}{r(r!)^24^r}\right).
\eqno(2.34)$$}\psp

{\bf Remark 2.3}. Applying the transformation (1.5), we obtain the
following non-steady solution:
\begin{eqnarray*}\hspace{1cm}H&=&(\al'-1)y+\be+b\sum_{i=0}^{\infty}\frac{k^i((x+\al)^2+y^2)^i}{(i!)^24^i}
\\ & &+c\sum_{j=0}^{\infty}((x+\al)^2+y^2)^j
\left(\frac{k^j\ln((x+\al)^2+y^2)}{(j!)^24^j}-2\sum_{r=0}^j\frac{k^r}{r(r!)^24^r}\right),
\hspace{1.9cm}(2.35)\end{eqnarray*}
 where $\al$ and $\be$ are
arbitrary functions of $t$.

\section{Solutions of Nonlinear MHD Equations}

In this section, we will find multiple parameter function exact
solutions of the nonlinear MHD equations (1.2) and (1.3).

 We first assume
$$\vf=\sgm_txy+ fx+g
y+h,\qquad \psi=\sgm_zxy+\lmd x+\mu y+\rho,\eqno(3.1)$$ where
$f,g,h,\sgm,\lmd,\mu$ and $\rho$ are functions in $t,z$. Then (1.3)
naturally holds. Moreover, (1.2) becomes $$(\lmd_t-f_z)
x+(\mu_t-g_z) y+\rho_t-h_z+(\sgm_ty+f)(\sgm_zx+\mu)
-(\sgm_tx+g)(\sgm_zy+\lmd)=0,\eqno(3.2)$$ equivalently,
$$\lmd_t-f_z+\sgm_zf-\sgm_t\lmd=0,\eqno(3.3)$$
$$\mu_t-g_z+\sgm_t\mu-\sgm_zg=0,\eqno(3.4)$$
$$\rho_t-h_z+f\mu-g\lmd=0.\eqno(3.5)$$
Solving (3.3) and (3.4), we get
$$f=\vt_te^\sgm,\qquad g=\tau_te^{-\sgm},\qquad
\lmd=\vt_ze^\sgm,\qquad\mu=\tau_ze^{-\sgm},\eqno(3.6)$$ where
$\vt$ and $\tau$ are arbitrary functions in $t,z$. Moreover, (3.5)
becomes
$$\rho_t-h_z+\vt_t\tau_z-\vt_z\tau_t=0.\eqno(3.7)$$
Thus
$$h=\vt_t\tau,\qquad \rho=\vt_z\tau\eqno(3.8)$$
modulo the transformation of  type $T_{0,\nu}$ in (1.9). So we have
the following simple conclusion.\psp

{\bf Proposition 3.1}. {\it Let $\sgm,\vt$ and $\tau$ be functions
of $t,z$. We get the following solution of the nonlinear MHD
equations (1.2) and (1.3):
$$\vf=\sgm_txy+ \vt_te^\sgm x+\tau_te^{-\sgm}
y+\vt_t\tau,\eqno(3.9)$$ $$\psi=\sgm_zxy+ \vt_ze^\sgm
x+\tau_ze^{-\sgm} y+\vt_z\tau.\eqno(3.10)$$} \psp

The main objective in this section is to find more sophisticated
exact solutions of the equations (1.2) and (1.3). Let $\Im$ and
$\ves$ be functions in $t,z$. Set
$$\varpi=\Im x+\ves y.
\eqno(3.11)$$ Suppose
$$\vf=\zeta(t,z,\varpi)+\sgm_txy,\qquad
\psi=\eta(t,z,\varpi)+\sgm_zxy,\eqno(3.12)$$ where $\zeta$ and
$\eta$ are functions in $t,z,\varpi$ to be determined, and $\sgm$
is a function of $t,z$. Then (1.2) becomes
$$\eta_t-\zeta_z+[\Im_tx+\ves_ty-\sgm_t(\Im x- \ves
y)] \eta_\varpi-[\Im_zx+\ves_zy-\sgm_z(\Im x- \ves
y)]\zeta_\varpi=0\eqno(3.13)$$ and (1.3) becomes
\begin{eqnarray*}& &(\Im^2+\ves^2)_t\zeta_{\varpi\varpi}-
(\Im^2+\ves^2)_z\eta_{\varpi\varpi}+(\Im^2+\ves^2)\{
\zeta_{t\varpi\varpi}+[\Im_tx+\ves_t y-\sgm_t(\Im x-\ves
y)]\zeta_{\varpi\varpi\varpi}\\ &
&-\eta_{z\varpi\varpi}-[\Im_zx+\ves_zy-\sgm_z(\Im x-\ves
y)]\eta_{\varpi\varpi\varpi}\}=0.\hspace{6.2cm}(3.14)\end{eqnarray*}

In order to solve the above system of partial differential
equation, we assume
$$\Im_sx+\ves_sy-\sgm_s(\Im x-\ves y)=0,\qquad s=t,z,\eqno(3.15)$$
equivalently,
$$\Im_s-\sgm_s\Im=0,\qquad \ves_s+\sgm_s\ves=0,\qquad s=t,z.\eqno(3.16)$$
So
$$\Im=b e^{\sgm},\qquad \ves=ce^{-\sgm},\qquad
b,c\in\mbb{R}.\eqno(3.17)$$ Moreover, (3.13) becomes
$$\eta_t-\zeta_z=0.
\eqno(3.18)$$ Hence
$$\zeta=F_t(t,z,\varpi),\qquad\eta=F_z(t,z,\varpi)
\eqno(3.19)$$ for some three variable function $F$. Now
 (3.14) becomes:
$$(\Im^2+\ves^2)_tF_{t\varpi\varpi}-
(\Im^2+\ves^2)_zF_{z\varpi\varpi}+(\Im^2+\ves^2)(F_{tt\varpi\varpi}
-F_{zz\varpi\varpi})=0.\eqno(3.20)$$ Modulo the transformation in
(1.9), we have
$$F=\hat F+\lmd\varpi\eqno(3.21)$$ for a function $\lmd$ of $t,z$
and a function $\hat F$ of $t,z,\varpi$ such that
$$(\Im^2+\ves^2)_t\hat F_t-
(\Im^2+\ves^2)_z\hat F_z+(\Im^2+\ves^2)(\hat F_{tt} -\hat
F_{zz})=0.\eqno(3.22)$$ Rewrite (3.22) as
$$((\Im^2+\ves^2)\hat F_t)_t-((\Im^2+\ves^2)\hat
F_z)_z=0.\eqno(3.23)$$

In order to find exact solutions of the above equation, we take
the following special cases of $\Im^2+\ves^2$. Without loss of
generality, we assume $b=1$ in (3.17).

\psp

{\it Case 1}. $\Im^2+\ves^2=e^{a_1t+a_2z}$ with
$a_1,a_2\in\mbb{R}$.\psp

Our assumptions says
$$e^{2\sgm}+c^2e^{-2\sgm}=e^{a_1t+a_2z},\eqno(3.24)$$
equivalently,
$$(e^{2\sgm})^2-e^{a_1t+a_2z}e^{2\sgm}+c=0.\eqno(3.25)$$
So
$$e^{2\sgm}=\frac{e^{a_1t+a_2z}+\sqrt{e^{2(a_1t+a_2z)}-4c^2}}{2}
\eqno(3.26)$$ or
$$e^{2\sgm}=\frac{e^{a_1t+a_2z}-\sqrt{e^{2(a_1t+a_2z)}-4c^2}}{2}.
\eqno(3.27)$$ Hence
$$\sgm=\frac{1}{2}
\left(\ln\left(e^{a_1t+a_2z}+\sqrt{e^{2(a_1t+a_2z)}-4c^2}\right)-\ln
2\right)\eqno(3.28)$$ or
$$\sgm=\frac{1}{2}
\left(\ln\left(e^{a_1t+a_2z}-\sqrt{e^{2(a_1t+a_2z)}-4c^2}\right)-\ln
2\right).\eqno(3.29)$$

Now (3.23) is equivalent to
$$\hat F_{tt}+a_1\hat F_t=\hat F_{zz}+a_2\hat
F_z.\eqno(3.30)$$ For $0\neq a\in\mbb{R}$, we denote
$${\cal D}(a,s)=a\ptl_s+\ptl_s^2\eqno(3.31)$$
and
$$\xi_{1,0}(a,s)=1,\qquad
\xi_{1,i}(a,s)=\sum_{r=0}^{i-1}\frac{(-1)^rs^{i-r}}{(i-r)!a^{i+r}},
\eqno(3.32)$$
$$\xi_{2,0}(a,s)=e^{-as},\qquad
\xi_{2,i}(a,s)=(-1)^ie^{-as}\sum_{r=0}^{i-1}\frac{s^{i-r}}{(i-r)!a^{i+r}},
\eqno(3.33)$$ for $0<i\in\mbb{Z}$. Moreover, we let
$${\cal
D}(0,s)=\ptl_s^2,\qquad\xi_{1,i}(0,s)=\frac{s^{2i}}{(2s)!},\qquad
\xi_{2,i}(0,s)=\frac{s^{2i+1}}{(2i+1)!}\eqno(3.34)$$ for $0\leq
i\in\mbb{Z}$. Then
$${\cal D}(a,s)(\xi_{\es,0}(a,s))=0,\qquad{\cal
D}(a,s)(\xi_{\es,i}(a,s))=\xi_{\es,i-1}(a,s),\qquad\es=1,2,\eqno(3.35)$$
 for
$0<i\in\mbb{Z}$. Given one-variable functions
$$\{\al_{\es,i_\es}(\varpi),\be_{\es,j_\es}(\varpi)\mid
\es=1,2;\;i_\es=0,1,...,m_\es;\;j_\es=0,1,...,n_\es\},\eqno(3.36)$$
 we have
the following solution of (3.30):
$$\hat F=\sum_{\es=1,2}[\sum_{i=0}^{m_\es}\al_{\es,i}(\varpi)\xi_{\es,i}(a_1,t)
\xi_{1,m_\es-i}(a_2,z)+\sum_{j=0}^{n_\es}\be_{\es,j}(\varpi)\xi_{\es,j}(a_1,t)
\xi_{2,n_\es-j}(a_2,z)].\eqno(3.37)$$ \pse

{\it Case 2}. {\it Case 1}. $\Im^2+\ves^2=e^{a_1t}z^{a_2}$ with
$a_1,a_2\in\mbb{R}$ such that $a_2\not\in\{0,\mbb{Z}+1/2\}$.\psp

As (3.24)-(3.28), we have:
$$\sgm=\frac{1}{2}
\left(\ln\left(e^{a_1t}z^{a_2}+\sqrt{e^{2a_1t}z^{2a_2}-4c^2}\right)-\ln
2\right)\eqno(3.38)$$ or
$$\sgm=\frac{1}{2}
\left(\ln\left(e^{a_1t}z^{a_2}-\sqrt{e^{2a_1t}z^{2a_2}-4c^2}\right)-\ln
2\right).\eqno(3.39)$$ In this case, (3.23) is equivalent to
$$\hat F_{tt}+a_1\hat F_t=\hat F_{zz}+a_2z^{-1}\hat
F_z.\eqno(3.40)$$ For $a\in\mbb{R}\setminus\{0,\mbb{Z}+1/2\}$,
$$\zeta_{1,0}(a,s)=1,\qquad\zeta_{1,i}(a,s)=\frac{s^{2i}}{2^ii!
\prod_{r=1}^i(a+2r-1)},\eqno(3.41)$$
$$\zeta_{2,0}(a,s)=z^{1-a},\qquad\zeta_{2,i}(a,s)=\frac{s^{2i+1-a}}{2^ii!
\prod_{r=1}^i(2r+1-a)}\eqno(3.42)$$ for $0<i\in\mbb{Z}$. Denote
$$\hat{\cal D}(a,s)=\ptl_s^2+\frac{a}{s}\ptl_s.\eqno(3.43)$$
Then $$\hat{\cal D}(a,s)(\zeta_{\es,0}(a,s))=0,\qquad\hat{\cal
D}(a,s)(\zeta_{\es,i}(a,s))=\zeta_{\es,i-1}(a,s),\qquad\es=1,2,
\eqno(3.44)$$
 for
$0<i\in\mbb{Z}$. Given the one-variable functions in (3.36), we
have the following solution of (3.40):
$$\hat F=\sum_{\es=1,2}[\sum_{i=0}^{m_\es}\al_{\es,i}(\varpi)\xi_{\es,i}(a_1,t)
\zeta_{1,m_\es-i}(a_2,z)+\sum_{j=0}^{n_\es}\be_{\es,j}(\varpi)\xi_{\es,j}(a_1,t)
\zeta_{2,n_\es-j}(a_2,z)].\eqno(3.45)$$ \pse

{\it Case 3}. {\it Case 1}. $\Im^2+\ves^2=t^{a_1}z^{a_2}$ with
$a_1,a_2\in\mbb{R}\setminus\{0,\mbb{Z}+1/2\}$.\psp

As (3.24)-(3.28), we have:
$$\sgm=\frac{1}{2}
\left(\ln\left(t^{a_1}z^{a_2}+\sqrt{t^{2a_1}z^{2a_2}-4c^2}\right)-\ln
2\right)\eqno(3.46)$$ or
$$\sgm=\frac{1}{2}
\left(\ln\left(t^{a_1}z^{a_2}-\sqrt{t^{2a_1}z^{2a_2}-4c^2}\right)-\ln
2\right).\eqno(3.47)$$ In this case, (3.23) is equivalent to
$$\hat F_{tt}+a_1t^{-1}\hat F_t=\hat F_{zz}+a_2z^{-1}\hat
F_z.\eqno(3.48)$$ Given the one-variable functions in (3.36), we
have the following solution of (3.48):
$$\hat F=\sum_{\es=1,2}[\sum_{i=0}^{m_\es}\al_{\es,i}(\varpi)\zeta_{\es,i}(a_1,t)
\zeta_{1,m_\es-i}(a_2,z)+\sum_{j=0}^{n_\es}\be_{\es,j}(\varpi)\zeta_{\es,j}(a_1,t)
\zeta_{2,n_\es-j}(a_2,z)].\eqno(3.49)$$

In summary, we have the following theorem:\psp

{\bf Theorem 3.2}. {\it Let $\lmd$ be any function of $t,z$ and
let $c$ be arbitrary real constant. Suppose the one-variable
functions in (3.36) are given. We have the following solution of
the nonlinear MHD equations (1.2) and (1.3):
$$\vf=\sgm_txy+\lmd_t(e^\sgm x+ce^{-\sgm}y)+\hat F_t(t,z,e^\sgm
x+ce^{-\sgm}),\eqno(3.50)$$
$$\psi=\sgm_zxy+\lmd_z(e^\sgm x+ce^{-\sgm}y)+\hat F_z(t,z,e^\sgm
x+ce^{-\sgm}),\eqno(3.51)$$ where (1) $\sgm$ is given in (3.28) or
(3.29) with $a_1,a_2\in\mbb{R}$ and $\hat F(t,z,\varpi)$ is given in
(3.37) via (3.32)-(3.34); (2) $\sgm$ is given in (3.38) or (3.39)
with $a_1,a_2\in\mbb{R}$ such that $a_2\not\in\{0,\mbb{Z}+1/2\}$ and
$\hat F(t,z,\varpi)$ is given in (3.44) via (3.32)-(3.34) and
(3.41)-(3.42); (3) $\sgm$ is given in (3.46) or (3.47) with
$a_1,a_2\in\mbb{R}\setminus\in\{0,\mbb{Z}+1/2\}$ and $\hat
F(t,z,\varpi)$ is given in (3.49) via (3.41) and (3.42).}\psp

Finally, we have the following obvious results.\psp

{\bf Proposition 3.3}. {\it Let $F(w,\varpi)$ and $G(w,\varpi)$ be
any two-variable functions. We have the following solutions of the
nonlinear MHD equations (1.2) and (1.3):
$$(1)\qquad\vf=F_w(t+z,x)+G_w(t-z,x),\qquad\psi=F_w(t+z,x)-G_w(t-z,x);
\eqno(3.52)$$ (2)
$$\vf=F_w(t+z,x^2+y^2)+G_w(t-z,x^2+y^2),\eqno(3.53)$$
$$\psi= F_w(t+z,x^2+y^2)-G_w(t-z,x^2+y^2). \eqno(3.54)$$}
\pse

We remark that we would get more sophisticated solutions if we apply
the transformations (1.6)-(1.16) to our above solutions. For
instance, applying the transformations (1.11)-(1.16) to the solution
in  (3.52), we obtain the following solution of the nonlinear MHD
equations (1.2) and (1.3):
\begin{eqnarray*}\vf&=&F_w(t+z,(x+\sgm)\cos2\al+(y+\tau)\sin2\al)
+\al'((x+\sgm)^2+(y+\tau)^2)\\ & &
+\sgm_t(y+\tau)-\tau_tx+\lmd_t+G_w(t-z,(x+\sgm)\cos2\al+(y+\tau)
\sin2\al),\hspace{2.6cm}(3.55)
\end{eqnarray*}
\begin{eqnarray*}\psi&=&F_w(t+z,(x+\sgm)\cos2\al+(y+\tau)\sin2\al)
+\es\al'((x+\sgm)^2+(y+\tau)^2) \\ &
&+\sgm_t(y+\tau)-\tau_tx+\lmd_t-G_w(t-z,(x+\sgm)\cos2\al+(y+\tau)\sin2\al)\hspace{2.8cm}(3.56)
\end{eqnarray*}
with $\es=\pm 1$, where $\al=\al(t+\es z)$ is an arbitrary
one-variable function, and $\sgm,\tau,\lmd$ are any functions in
$t,z$.

 \vspace{0.5cm}

\noindent{\Large \bf References} \vspace{0.2cm}

\begin{description}

\item[{[B]}] A. G. Bershadskii, Energy conservation law and
uniqueness of the classical solutions of nonlinear MHD problems,
{\it Magnetohydrodynamics} {\bf 2} (1980), 78-80.

\item[{[I]}] N. H. Ibragimov, {\it Lie Group Analysis of
Differential Equations}, Volume 2, CRC Handbook, CRC Press, 1995.

\item[{[LRT]}] C. C. Lin, E. Reissner and H. S. Tsien, On
two-dimensional non-steady motion of a slender body in a
compressible fluid, {\it J. Math. Phys.} {\bf 27} (1948), no. 3,
220.

\item[{[Ka1]}] V. I. Katkov, One class of exact solutions of the
geopotential forecast equation, {\it Izvestiya Akad. Nauk
S.S.S.R., Fizica Atmosferi i Okcana} {\bf 1} (1965), 1088.

\item[{[Ka2]}] V. I. Katkov, Exact solutions of the geopotential
forecast equation, {\it Izvestiya Akad. Nauk S.S.S.R., Fizica
Atmosferi i Okcana} {\bf 2} (1966), 1193.

\item[{[Ki]}] T. W. Kibel', {\it Introduction to Hydrodynamical
Methods of Short Term Weather Forecast}, Gosteoretizdat, Moscow,
1954.

\item[{[M]}] E. V. Mamontov, On the theory of nonstationary
transonic flows, {\it Dokl. Acad. Nauk SSSR} {\bf 185} (1969), no.
3, 538.

\item[{[O]}] L. V. Ovsiannikov, {\it Group Analysis of
Differential Equations}, Academic Press, New York, 1982.

\item[{[P]}] Shih-I Pai, {\it Magnetogasdynamic and Plasma
Physics}, Springer-Verlag, Wien, 1962.

\item[{[Sa]}] A. V. Samokhin, Nonlinear MHD equations: symmetries,
solutions, and conservation laws, {\it Dokl. Akad, Nauk S.S.S.R.}
{\bf 285} (1985), no. 5, 1101-1106.

\item[{[Sy]}] S. Syono, Various simple solutions of the barotropic
vorticity equation, in {\it Vortex, Collected papers of the
numerical weather prediction group in Tokyo} (1958), 3.

\item[{[X1]}] X. Xu,  Stable-Range approach to the equation of
nonstationary transonic gas flows, {\it Quart. Appl. Math.} {\bf 65}
(2007), 529-547.

\item[{[X2]}] X. Xu, New algebraic approaches to
 classical boundary layer problems, {\it arXiv:0706.1864}.

\item[{[X3]}] X. Xu, Asymmetric and moving-frame  approaches to
Navier-Stokes equations, {\it  Quart. Appl. Math.}, in press,
arXiv:0706.1861.

\end{description}

\end{document}